\documentclass[useAMS,senatbib,usegraphicx,letterpaper]{mn2e}
\usepackage{graphics}
\usepackage{graphicx}
\usepackage{psfig}
\usepackage{float}
\usepackage{psfrag}
\usepackage{amsmath}
\usepackage{amssymb}
\usepackage{ulem}
\usepackage{ctable}
\usepackage{txfonts}
\usepackage{enumerate}

\newcommand{\ysz}{Y_{\mathrm{SZ}}}
\newcommand{\plk}{{\it Planck~}}
\newcommand{\vir}{_{\mathrm{vir}}}

\title[SZ scaling of cluster radio haloes]{A Sunyaev-Zel'dovich take on cluster radio haloes -- I. Global scaling and bi-modality using {\it Planck} data}

\author[K. Basu]{Kaustuv Basu\thanks{E-mail:
kbasu@astro.uni-bonn.de} \\
Argelander Institut f\"ur Astronomie, Universit\"at Bonn, Auf dem H\"ugel 71, 53121 Bonn, Germany
}

\begin{document}

\date{Accepted 2012 January 4. Received 2012 January 3; in original form 2011 September 15}

\pagerange{\pageref{firstpage}--\pageref{lastpage}} \pubyear{2011}

\maketitle

\label{firstpage}

\begin{abstract}
\noindent
Giant radio haloes in galaxy clusters are the primary evidence for the existence of relativistic particles (cosmic rays) and magnetic fields over Mpc scales. Observational tests for the different theoretical models explaining their powering mechanism have so far been obtained through X-ray selection of clusters, e.g. by comparing cluster X-ray luminosities with radio halo power. Here we present the first global scaling relations between radio halo power and integrated Sunyaev-Zel'dovich (SZ) effect measurements, using the {\it Planck} all-sky cluster catalog and published radio data. The correlation agrees well with previous scaling measurements  based on X-ray data, and offers a more direct probe into the mass dependence inside radio haloes. However, we find no strong indication for a bi-modal cluster population split between radio halo and radio quiet objects. We discuss the possible causes for this apparent lack of bi-modality, and compare the observed slope of the radio-SZ correlation with competing theoretical models of radio halo origin.

\end{abstract}


\section{Introduction}

The bulk of the baryonic mass in galaxy clusters exists in the form of a low density, diffused ionized gas filling up the space between cluster galaxies. This hot intra-cluster medium (ICM, $T \sim 2-10$ keV) emits in the X-rays through thermal Bremsstrahlung emission, and is also observable in the millimeter/sub-millimeter wavelengths through the Sunyaev-Zel'dovich (SZ) effect, which is a distortion in the intensity of the Cosmic Microwave Background (CMB) radiation caused by the same thermal component (Sunyaev \& Zel'dovich 1980). Together, these two observables are central to the use of galaxy clusters as cosmological probes.

The ICM is also host to a large population of ultra-relativistic particles (cosmic rays) and magnetic fields, seen primarily through radio observations. The most spectacular evidence for this non-thermal population comes from observations of giant radio haloes, which are diffuse sources of radio synchrotron emission extending over $\sim 1$ Mpc scales.  The haloes are not associated to any particular cluster galaxy, and are morphologically distinct from radio mini-haloes (residing in cluster cool cores), radio relics (formed at the edge of a merger shock) and radio lobes associated with active galactic nuclei. The similarity of their morphology with the ICM suggests a correspondence between their powering mechanism and the total cluster mass (e.g. Liang et al. 2000).  They are relatively rare and are found mostly in clusters showing evidence of ongoing mergers. As such, they can prove to be essential in understanding cluster merger dynamics and associated heating processes in the ICM (see e.g. review by Ferrari et al. 2008).

Despite their importance, the powering mechanism of these giant radio haloes remains uncertain. There are two models for particle acceleration in a radio halo volume: the hadronic model which uses collisions between cosmic-ray protons and thermal protons for generating relativistic electrons (Dennison 1980), and the turbulence models where the electrons are re-accelerated through MHD turbulence in the ICM caused by cluster mergers (Brunetti et al. 2001, Petrosian 2001). The distinction between these two models is partly based upon the observed scaling between radio and X-ray power (the latter indicating the total cluster mass), and the fact that X-ray selection seems to indicate two distinct populations of clusters: the radio halo and ``radio quiet" ones (e.g Brunetti et al. 2007). However, recent discoveries of radio haloes in clusters with very low X-ray luminosities (Giovannini et al. 2011), and the lack of radio haloes in some mergers  (e.g. Russell et al. 2011) show that the X-ray selection may not be as clean as expected. These new observations and the underlying large scatter in the $L_{X}-P_{\mathrm{radio}}$ correlation suggest that a new observational window on the selection and mass estimation of clusters harboring radio haloes  can bring some much needed clarity.

One further reason for expecting a robust correlation between radio power and SZ is the timescale argument: the boost in the X-ray luminosity during mergers happens in a relatively short timescale, compared to the gas thermalization in a modified potential well producing a more gradual and moderate increase in the SZ signal (Poole et al. 2007, Wik et al. 2008). This should correspond better with the radio halo time scale ($\sim 1$ Gyr), derived from the spatial extent of the haloes. The integrated SZ signal is also a more robust indicator of cluster mass than the X-ray luminosity, irrespective of cluster dynamical state (e.g. Motl et al. 2005, Nagai 2006). Thus SZ-selection might be able to find radio haloes in late mergers and other massive systems which are left out in X-ray selection.
\medskip 

This letter presents the first radio-SZ correlation for clusters with radio haloes. The radio data is a compilation of published results, and the SZ measurements are taken from the \plk all-sky cluster catalog (Planck collaboration 2011).  All results are derived using the $\Lambda$CDM concordance cosmology with $\Omega_M = 0.26$, $\Omega_{\Lambda} = 0.74$ and $H_0 = 71$ km s$^{-1}$ Mpc$^{-1}$.  The quantity $\ysz$ is used throughout to denote the {\it intrinsic} Compton $Y$-parameter for a cluster, $Y d_A^2$, where $d_A$ is its angular diameter distance. 

\vspace*{-2mm}
\section{Radio \& SZ data sets}

We do not attempt to define a new comprehensive sample for this work, rather use a set of available cluster catalogs with radio halo detections and non-detections to probe the robustness of the radio-SZ scaling. 
Published radio error estimates often ignore systematic effects like flux loss in interferometric imaging and contribution from unresolved point sources, which in turn create an over-estimation of the intrinsic scatter.  
Since the present work is mainly concerned  with the mean slope of the radio-SZ scaling and not its dispersion, error underestimation   in the literature will not affect the results as long as the measurements are unbiased.
 
The radio catalogs can be divided into two groups: those with and without a listing of non-detections. A comprehensive  sample in the former category is by Giovannini et al. (2009, hereafter G09), presenting results at 1.4 GHz for $z<0.4$ clusters. Potentially problematic are its mixing of radio halos and mini halos, and not separating contributions from radio relics. More critically, the sizes of the radio haloes are approximated by the observed largest linear sizes (LLS), which is not a good approximation for radio halo diameter. To address the latter issue we use a smaller subsample by Cassano et al. (2007, hereafter C07), which provides a better measurement of radio halo sizes by averaging their minimum and maximum extensions. 

The most systematic study of radio halo non-detections in an X-ray selected sample is by Venturi et al. (2008), using GMRT observation at 610 GHz. However, this sample contains too few clusters which have \plk SZ measurements to obtain any robust correlation. We therefore use the compilation by Brunetti et al. (2009, hereafter B09), which lists GMRT results scaled to 1.4 GHz with other unambiguous halo detections. The non-detection upper limits were obtained by simulating fake radio halos in the GMRT data and scaling to 1.4 GHz by using $\alpha=1.3$, a typical spectral index for radio haloes. Our final sample is from Rudnick \& Lemmerman (2009, hereafter R09) who re-analyzed WENSS survey data at 327 MHz for an X-ray selected sample. The shallowness of WENSS data makes R09 ineffective in testing bi-modality, as the $3\sigma$ upper limits are not sufficiently below the detection level. Its use is mainly limited to testing possible changes in the scaling law at lower frequencies.

The Sunyaev-Zel'dovich effect measurements are taken from the Planck ESZ catalog (Planck collaboration 2011). This all-sky cluster catalog provides a list of 189 objects in the highest mass range out to a redshift $z\sim 0.6$, selected at $S/N > 6$ from the first year survey data. Out of these, 22 are either new cluster candidates or have no X-ray data. The remaining 167 clusters, spanning a redshift range $0.01 <z <0.55$, are cross-correlated against the radio catalogs. All radio halo clusters therefore have an $R_{500}$ estimate in the \plk catalog obtained from the $L_X-M_{500}$ relation. This is used to model the pressure profile for each cluster when scaling their integrated SZ signal, $\ysz$, between different radii.

Proper regression analysis between the radio and SZ observables is fundamental to this work. We must allow for measurement errors and intrinsic scatter in both observables, which makes the regression analysis non-trivial (see Kelly 2007). We use the publicly available \textsc{idl} code by Kelly to perform the regression analysis using a Bayesian approach. An important advantages of this method is the provision for including non-detections. 

\begin{table}
\caption{Regression coefficients for the scaling relation $\log(P_{\nu}) = A + B ~\log(\ysz)$. 
The term {\it global} implies correlation with the total SZ signal, as opposed to that scaled inside the halo radius. }
\label{regtable}
\centering
\begin{tabular}{l l c c}
\hline
Sample & sub-category  & ~~B (slope)~~ & ~~A (norm.)~~ \\
\hline\hline 
\small
 G09 & global & $1.84\pm 0.38$ & $31.3\pm1.4$ \\
        & inside LLS & $0.95\pm 0.14$ & $28.8\pm 0.5$ \\

 C07 & global & $1.88\pm 0.24$ & $31.4\pm0.8$ \\
        & inside $R_{H}$ & $1.17\pm 0.18$ & $29.7\pm0.8$ \\
 
 B09 & global, haloes only & $2.03\pm 0.28$ &  $32.1\pm 1.0$ \\
        & $+$ non-detections &  $2.41\pm 0.44$ &  $33.4\pm 1.6$ \\

R09 & global, haloes only & $0.81\pm 0.36$ &  $28.1\pm 1.4$ \\
        & $+$ non-detections & $1.38\pm 0.43$ & $29.8\pm 1.8$ \\
\hline
 \end{tabular}
 \begin{minipage}[b]{0.92\columnwidth}
 \centering
 \vspace{1mm}
 Samples: G09=Giovannini et al. 2009; C07=Cassano et al. 2007; B09=Brunetti et al. 2009; 
 R09=Rudnick \& Lemmerman 2009.
 \end{minipage}
 \end{table}

   \begin{figure*}
   \includegraphics[width=0.92\columnwidth]{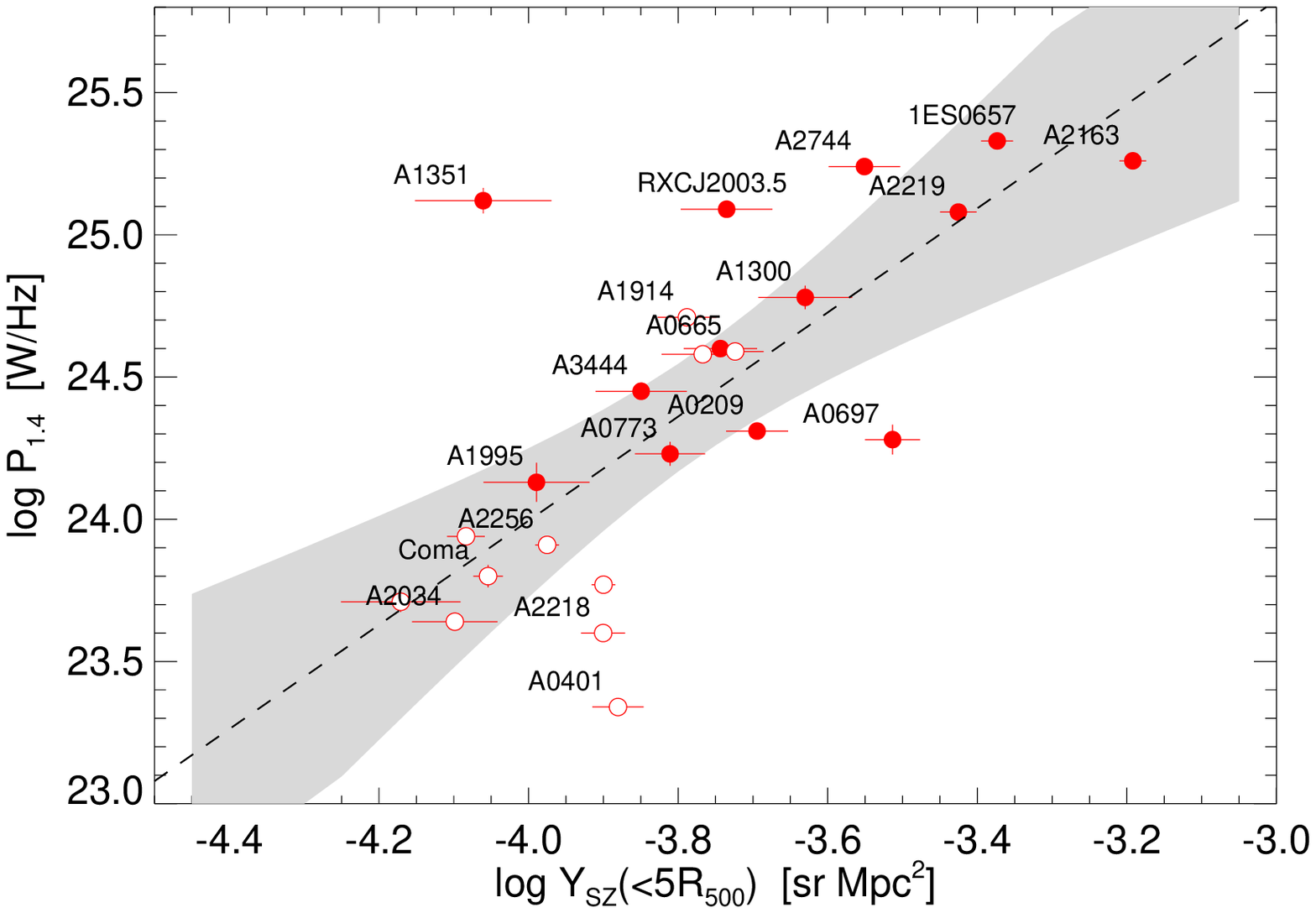}
   \hspace{4mm}
   \includegraphics[width=0.92\columnwidth]{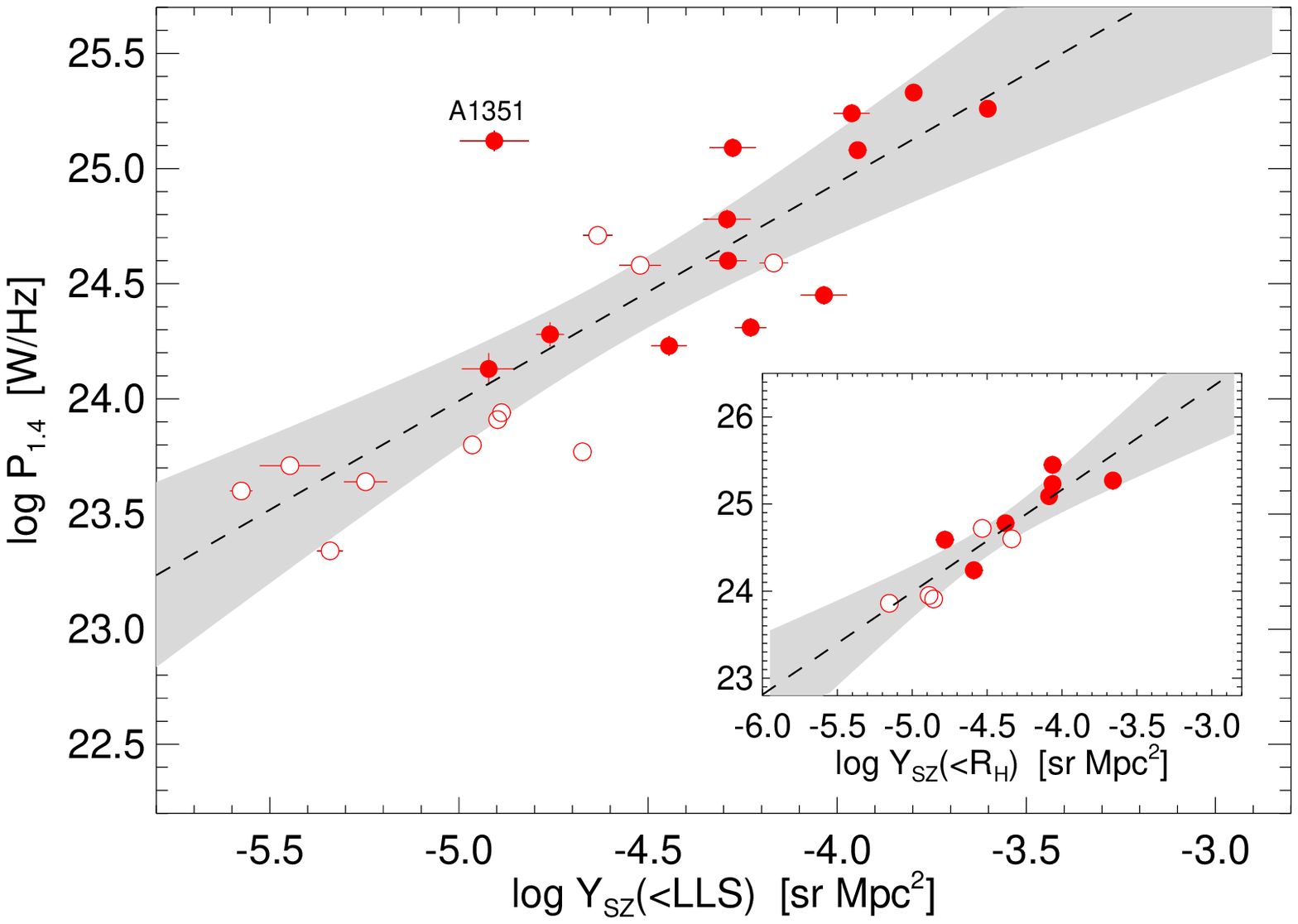}
    \caption{Radio-SZ correlation for {\it Planck} detected clusters with radio haloes. {\it Left --} Result from the G09 sample, correlating radio halo power against the total $\ysz$ signal inside $5R_{500}$. Filled symbols correspond to clusters at $z>0.2$, and open symbols are for lower redshifts. The shaded area marks the $2\sigma$ or 95\% confidence region  (some name labels are omitted for clarity). {\it Right --} The same G09 clusters after scaling the total SZ signal to inside the haloes' largest linear dimensions, resulting in a much flatter slope with reduced scatter. ({\it Inset}) Result from the C07 sample, with scaled SZ signal inside their quoted radio halo radius, $R_H$. Mean slope is $1.17\pm 0.18$, which is used for comparison with theoretical models.}
    \label{GCsamp}
   \end{figure*}

\vspace*{-2mm}
\section{Results}

\subsection{Radio-SZ scaling}

The radio-SZ scaling relation is obtained by performing linear regression in log-space: $\log(P_{\nu}) = A + B\log(\ysz)$. 
The normalization $A$ and slope $B$ are obtained from the Markov Chains, as well as the intrinsic scatter $\sigma_{\log P|\log Y}$. A summary of the results are given in Table 1.  The first correlation example is from the G09 sample, comparing the radio power with the {\it global} SZ signal (Fig. \ref{GCsamp} {\it left}).  Out of 32 objects in this sample 24 have {\it Planck} counterparts. The mean slope is $1.84\pm 0.38$.  There is a lot of scatter in this correlation, with mean  scatter 0.45 dex, i.e. roughly a factor $\sim 2.8$. Much of this scatter is driven by the low-redshift objects which are under-luminous (e.g. A401 and A754). This potentially indicates a systematic bias in their total flux and size measurements with interferometers.  Only A1351 stands out as overtly radio luminous for its mass, although later revisions of its radio power (Giacintucci et al. 2009) moves it closer to the mean value.

The \plk catalog provides the integrated $Y$ parameter within radius $5 R_{500}$, obtained from a matched filtering algorithm assuming a universal gas pressure profile (see Planck collaboration 2011).  At this radius, $Y_{5R_{500}}^{\mathrm{cyl}} \approx Y_{5R_{500}}^{\mathrm{sph}}$. This  is nearly 3 times the cluster virial radius, and much larger than the extent of the radio emitting regions. Therefore, a tighter correlation can be expected if this {\it global} SZ signal is scaled down to that inside the radio halo volume. We do this conversion by assuming the universal pressure profile of Arnaud et al. (2010), as also used by the {\it Planck} team. In particular, the best fit profile for mergers/disturbed clusters from the appendix of Arnaud et al. (2010) is used, but the difference is negligible if  the mean profile is used instead. 
This scaling of the SZ signal inside LLS changes the results significantly. Correlation between radio and SZ powers inside the halo volume becomes consistent with a linear relation, with mean slope $0.95\pm 0.14$ (Fig.\ref{GCsamp} {\it right}) and a reduced mean intrinsic scatter in radio (0.35 dex).

The largest linear sizes are in general not a good approximation for radio halo diameters, so the above analysis is repeated with the C07 sample using their revised measurement of halo radius, $R_H$. The slope for the global correlation with this sample is $1.88\pm 0.24$, whereas after scaling the SZ signal inside $R_H$ it becomes $1.17\pm 0.18$, with mean intrinsic scatter 0.28 dex (Fig.\ref{GCsamp} {\it inset}). Although this is statistically fully consistent with the scaled result of the G09 sample inside the LLS, we use this slightly super-linear correlation when making comparisons with theoretical models, due to the better definition of halo radius. We emphasize that from the current analysis using radio data from the literature, a linear correspondence between radio and SZ power is a fully valid result.

The R09 sample at 327 MHz indicates a flattening of the correlation slope at lower frequencies: the best fit value is $0.81\pm 0.36$ when considering the halo sample, with a scatter of only 0.21 dex. The large flux uncertainties (and correspondingly low intrinsic scatter) reflect on the shallowness of the WENSS data, which is more than an order of magnitude less sensitive compared to typical VLA measurements scaled to its frequency. However, the method used by R09 to detect haloes (and place upper limits) based on simulating sources in control regions safeguards against flux underestimation bias, which will otherwise occur in a visual inspection.  
There can be a residual bias from fluxes associated with small scale structures that are not recovered. 
If non-detection upper limits in R09 are included in the correlation, then the slope becomes $1.38\pm 0.43$, which is consistent at $1\sigma$ with the scaling result at 1.4 GHz.

   \begin{figure*}
   \includegraphics[width=0.92\columnwidth]{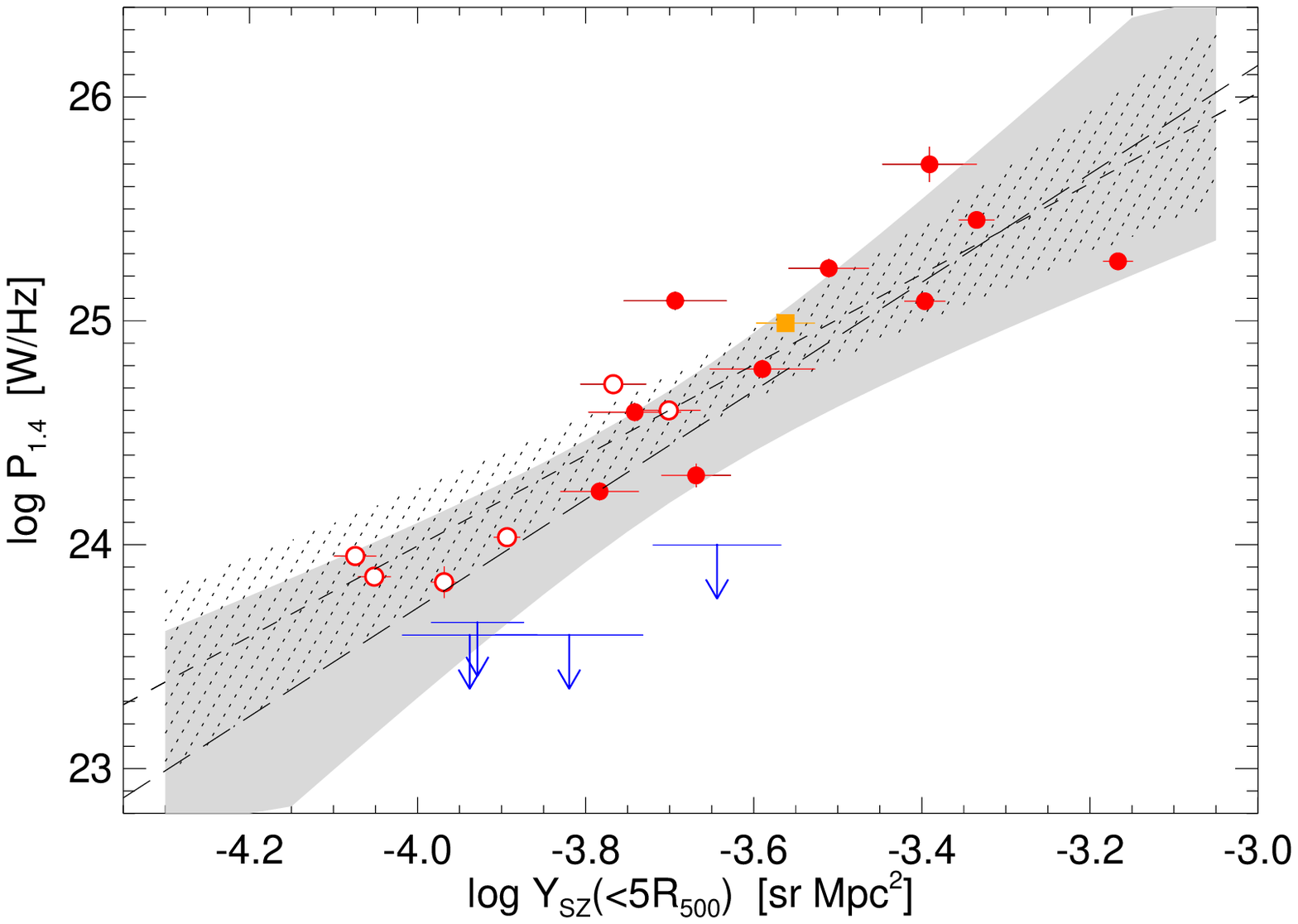}
   \hspace{4mm}
   \includegraphics[width=0.92\columnwidth]{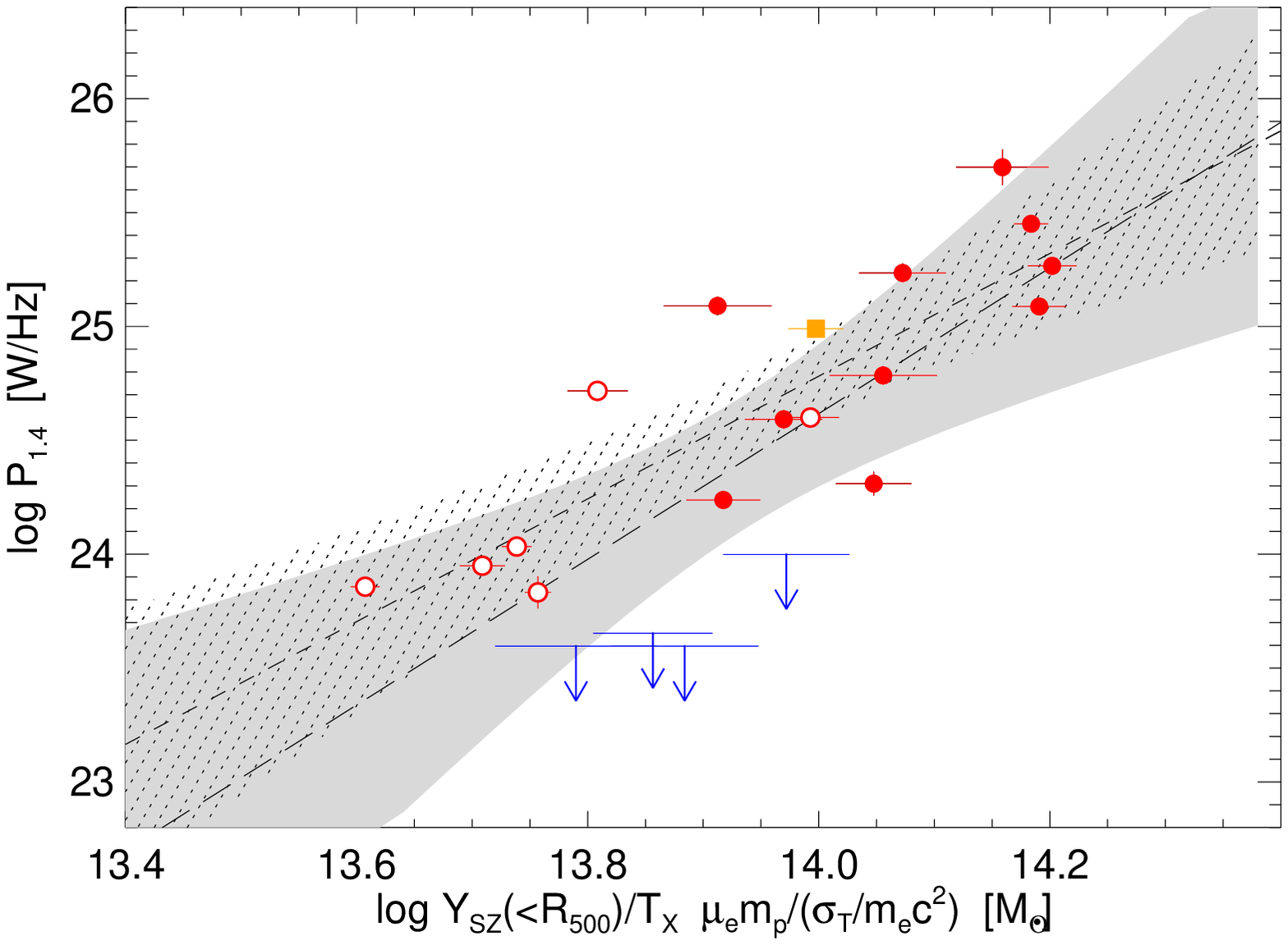}
      \caption{ Test of bi-modality with non-detections from the B09 sample.  
      {\it Left --} Correlation of radio power against the total SZ signal: filled symbols are for $z>0.2$ cluster haloes, and open symbols are for lower redshifts. All non-detections are at $z>0.2$ and are extrapolated from 610 MHz data. The only mini-halo (A2390) in the sample is marked by the orange square.  The short-dashed line corresponds to the fit for haloes only, and the long-dashed line when non-detections are included. Filled regions mark the 95\% confidence intervals. {\it Right --}  Correlation of radio halo power with the gas mass. Symbols and lines have the same meaning as in the left panel.
      }
      \label{Bsamp}
   \end{figure*}

\vspace*{-2mm}   
\subsection{Lack of strong bi-modality}

To test whether there are two distinct populations of clusters: those hosting powerful radio haloes and those without, we use the B09 sample. Actual flux measurements at 1.4 GHz are used for the clusters A209 and RXCJ1314 (Giovannini et al. 2009), instead of the extrapolated values from 610 MHz as given by B09. All non-detection upper limits are extrapolation of simulation results at 610 MHz using halo spectral index $\alpha=1.3$. 
Regression analysis for the two cases (with and without halo non-detections) yields slopes which are statistically consistent with each other, even though a bi-modal division appears to be emerging (Fig. \ref{Bsamp} {\it left}). Significantly, we do not find high-$\ysz$ objects with radio non-detections, as in the case of highly X-ray luminous ``radio quiet'' cool core clusters. But the small number of non-detections makes it difficult to conclude whether the bi-modality is weaker or non-existent. All non-detection upper limits lie below the 95\% confidence interval  of the halo-only correlation, suggesting clusters with upper limits are generally radio under-luminous. This can be partly redshift-driven, as the samples are not SZ-complete.

An alternative, although not independent, way to visualize this result is to correlate the radio power directly against  cluster gas mass. Rough estimates of  $M_{\mathrm{gas}}$ inside $R_{500}$ were obtained by dividing the Planck $\ysz$ values by the mean X-ray temperatures taken from the literature. The global $\ysz$ measurements of \plk are scaled to that inside $R_{500}$ using the universal pressure profile, as this is the radius within which X-ray temperatures are typically obtained. The result is similar to the $P_{1.4}-\ysz$ correlation, lacking a strong bi-modal division (Fig.\ref{Bsamp}, right panel). The scatter is increased by roughly 30\% (from 0.5 dex to 0.6 dex), in line with the expectation that $\ysz$ is a lower scatter mass proxy. The four non-detection clusters have generally lower $T_X$ values as compared to the halo detections (median $T_X$ is $7.3$ keV compared to $8.7$ keV). It should be made clear that by taking $T_X$ estimates from literature we ignore potential errors due to non-uniform radius for extracting $T_X$, systematic differences between XMM-Newton and Chandra measurements, etc. However, the mean slope for the $P_{1.4}-M_{\mathrm{gas}}$ scaling relation, $3.2\pm 0.7$ with the full sample, is consistent with the global mass scaling derived from the $Y-M$ relation (see \S\ref{Msection}), indicating that no additional biases are incurred while using this non-uniform selection of X-ray temperatures.

A tentative argument for a selection bias in  X-ray complete samples and the ensuing bi-modality can be given by comparing the relative frequency with which radio haloes and non-detections occur in the \plk catalog. In Venturi et al. (2008), GMRT data were obtained for a  complete X-ray selected sample, with 6 detection of radio haloes 
plus 20 non-detections. The \plk catalog contains 5 out of these 6 halo clusters, but only 4 out of 20 non-detection clusters. For the B09 sample this ratio is 16 out of 21 radio halo clusters and 4 out of 20 non-detections (the same non-detections as in the Venturi et al. sample). Since the \plk catalog should not have a significant bias towards mergers, this provides an indirect evidence for our hypothesis that being SZ-bright (hence massive) is a better indicator for clusters hosting radio haloes, as opposed to being X-ray luminous. Even though the R09 sample is too shallow to directly test bi-modality, it interestingly follows this same trend: \plk reports measurement of 12 out of 14 counterparts for haloes and other diffuse emissions, as opposed to only 15 out of 58 counterparts for non-detections.

\vspace*{-1mm}   
\section{Theoretical considerations}

\subsection{Mass scaling of the radio halo power}
\label{Msection}
The independent variable, $\ysz$, is defined as the integral of the total pressure in a spherical volume, and hence is proportional to the total gas mass:
\begin{equation}
\ysz \equiv Y d_A^2 \propto \int n_e T_e~ dV \propto M_{\mathrm{gas}} T_e = f_{\mathrm{gas}} M_{\mathrm{tot}} T_e.
\label{eq:szdef}
\end{equation}
Here $T_e$ is the mean gas temperature within the integration radius, and $f_{\mathrm{gas}}$ is the gas-to-mass ratio. 
Assuming hydrostatic equilibrium and isothermality, the temperature scales to the total mass as $T_e \propto M_{\mathrm{tot}}^{2/3} E(z)^{2/3}$ (e.g. Bryan \& Norman 1998), where $E(z)$ is the ratio of the Hubble parameter at redshift $z$ to its present value. Therefore, the scaling between the SZ observable and total mass is $\ysz E(z)^{-2/3} \propto f_{\mathrm{gas}} M_{\mathrm{tot}}^{5/3}$. Numerical simulations, analytical models and SZ observations indicate that this mass scaling is extremely robust, with  little scatter over a large range of cluster mass, dynamical state or other details of  cluster physics (e.g. Motl et al. 2005, Reid \& Spergel 2006, Andersson et al. 2011). 
We thus assume this scaling to be valid also {\it inside} a cluster at different radii, provided that the radius is sufficiently large to exclude  complex physics at clusters cores. The large halo sizes measured by C07 ($\bar{R}_H \sim 600$ kpc, typically of the same order as $R_{2500}$), ensures that they encompass a representative cluster volume. 
The $E(z)^{-2/3}$ factor for self-similar evolution changes the scaling results only marginally, well within the statistical errors. The gas mass fraction, $f_{\mathrm{gas}}$, has a weak dependence on cluster mass: $f_{\mathrm{gas}} \propto M_{\mathrm{500}}^{~0.14}$ (Bonamente et al. 2008, Sun et al. 2009). Assuming the same mass dependence of $f_{\mathrm{gas}}$ for all radii, we therefore obtain 
\begin{equation}
P_{1.4} \propto M_H^{~~2.1\pm 0.3} \propto M\vir^{~~3.4 \pm 0.4}.
\label{eq:mscale}
\end{equation}
In the above, $M_H$ is the total mass inside radio haloes, and $M\vir$ is the cluster virial mass which scales linearly with $M_{\mathrm{tot}}(<5R_{500})$. The scaling index inside haloes is in good agreement with previous X-ray hydrostatic mass estimates (e.g. Cassano et al. 2007). The global scaling with total cluster mass can be a useful parameter for estimating radio halo statistics, particularly in simulations.

The radio halo sizes are known to scale non-linearly with cluster radius, in a break from self-similarity (Kempner \& Sarazin 2001, Cassano et al. 2007). 
Indeed, using the X-ray derived $R_{500}$ measurements from the \plk catalog, we obtain the empirical relation $R_H \propto R_{500}^{\ 3.1 \pm 0.2}$ with the C07 sample, consistent with the estimate by C07 using  $L_X-M\vir$ scaling relation ($R_H \propto R\vir^{2.6\pm 0.5}$).  A consequence of this rapid increase in radius is a drop of the mean gas density inside haloes with increasing halo mass. Our observed scaling between the halo radius and  scaled SZ signal, $R_H \propto Y_H^{\ 0.31\pm 0.03}$, implies that the mean gas density ($\bar{n}_H$)  scales down as roughly $\bar{n}_H \propto T_e^{-0.9}$, or assuming thermal equilibrium inside haloes, as $\bar{n}_H \propto M_H^{\ -0.6}$. This brings the observed non self-similar scaling between $R_H$ and $R\vir$ in  conformity with the mass scaling in Eq.(\ref{eq:mscale}). It is worth mentioning at this point that radio halo size measurements with insufficient S/N will tend to show a steeper dependence on luminosity than the true scaling, since only the bright central regions will be picked up.

\vspace*{-2mm}
\subsection{Comparison with radio/X-ray scaling}
There is some confusion in the literature about the exact power in the X-ray/radio scaling: reported values using the luminosity in  the soft X-ray band range between $P_{\nu} \propto L_{X[0.1-2.4]}^{1.6-2}$ (Brunetti et al. 2007, Kushnir et al. 2009).  Using the regression method used in this work we find the scaling index in the middle of this range, e.g. from the B09 sample the mean slope for $\log P_{\nu} - \log L_{X[0.1-2.4]}$ correlation is $1.80\pm 0.21$, with mean intrinsic scatter 0.3 dex. The mass-luminosity relation for the X-ray soft band is well-established observationally. We use the result given by Zhang et al. (2011) for disturbed clusters in the HIFLUGCS sample: $L_{X[0.5-2]} \propto [M_{\mathrm{gas}, R_{500}} E(z)]^{1.16\pm 0.04}$,  where the luminosities are core corrected. This combined with the weak mass dependence of $f_{\mathrm{gas}}$ produces a mass scaling of radio power as $P_{1.4} \propto M_{500}^{\ 2.4}$, which is much shallower than the virial mass scaling obtained in Eq.(\ref{eq:mscale}) but roughly consistent with the halo mass power law. This indicates that the global X-ray emission acts as a relatively good proxy for radio halo masses due to its peaked profile, as most of the X-ray flux comes from within a radius that is $\lesssim R_H$.

\vspace*{-2mm}
\subsection{Expectations from theoretical models}
The hadronic model for radio synchrotron emission postulates that electrons at ultra-relativistic energies are produced in the ICM by $p$-$p$ collisions between cosmic ray protons and thermal protons (see review by Ensslin et al. 2011 and references therein). For estimating the scaling relation between radio halo power and cluster mass, we follow the formulation by Kushnir et al. (2009). In this model, the total radio power is the volume integral of the cosmic ray energy density ($\epsilon_{\mathrm{CR}} = X_{\mathrm{CR}} ~n ~k T_e$) and the hadronic interaction rate ($\tau_{\mathrm{pp}}^{-1} \sim n ~\sigma_{\mathrm{pp}}$):
\begin{equation}
P_{\nu}  = \int \tau_{\mathrm{pp}}^{-1} ~\epsilon_{\mathrm{CR}} ~dV 
             \sim X_{\mathrm{CR}} ~n^2 ~k T_e  ~\sigma_{\mathrm{pp}} ~f_B ~R_H^3.\\
\end{equation}
Here $X_{\mathrm{CR}}$ is the ratio between cosmic ray pressure and thermal pressure, $n$ is the gas density, $\sigma_{\mathrm{pp}}$ is the $p$-$p$ collision cross-section, $f_B$ is the volume filling factor for magnetic fields, and $R_H$ is the halo radius. In the second step we have assumed the magnetic field energy density to be much larger than the CMB energy density, $B >> B_{\mathrm{CMB}} \approx 3.2(1+z)^2 \mu$G. 
Considering that $\ysz(<R_H) \sim n ~k T_e ~R_H^3$, we thus obtain:
$P_{\nu}  / \ysz \propto X_{\mathrm{CR}} ~f_B ~n ~\sigma_{\mathrm{pp}}$. 
Therefore, if the cosmic ray fraction and mean density do not depend on the halo mass, we recover the observed linear dependence between radio power and the SZ signal inside haloes. The latter assumption, however, is in conflict with our observed scaling of mean gas density (\S\ref{Msection}), which actually drops with increasing halo mass. Another potential problem is the assumption of strong magnetic fields, $B_H >> B_{\mathrm{CMB}}$, over the entire halo volume. A likely scenario with the hadronic model would therefore be to assume a more clumpy radio emission, where the regions contributing most of the radio power have constant densities and strong magnetic fields.

In the turbulent re-acceleration model a pre-existing population of electrons at lower energies gets re-accelerated by merger induced turbulence (see review by Ferrari et al. 2008 and references therein). The powering mechanism of radio haloes is complex, but for a simple estimate we can follow the formulation by Cassano \& Brunetti (2005) and Cassano et al. (2007). In their model, the energy injection rate from turbulence depends on the mean density and velocity dispersion inside the radio haloes; $\dot{\varepsilon}_t \propto n ~\sigma_H^2$, where $\sigma_H^2 \equiv GM_H/R_H$. The total power of a radio halo is then
\begin{equation}
P_{\nu}  \sim \int \dot{\varepsilon}_t ~(\Gamma_{\mathrm{rel}}/\Gamma_{\mathrm{th}}) 
	~dV \propto \dfrac{M_H~ \sigma_H^3}{{\cal F}(z, M_H, B_H)} ,
\label{eq:turb}
\end{equation}
where $\Gamma$ is the turbulence damping rate transferring energy to the particles, 
and the function ${\cal F}(z, M_H, B_H)$ is constant in the asymptotic limit of strong magnetic fields. Thus to the first approximation, Eq.(\ref{eq:turb}) implies $P_{\nu} \propto Y_H T_e^{1/2}$, in good agreement with the slightly super-linear slope inside haloes seen from the C07 sample ($P_{1.4} \propto Y_H^{\ 1.17 \pm 0.18}$). However, using the definition of $\sigma_H$ and the observed scaling between halo mass and radius, we find a mass dependence slightly shallower than observed: $P_{\nu} \propto M_H^{1.7-1.8}$, which is still consistent with Eq.(\ref{eq:mscale}). This may indicate a preference for more realistic field strengths,
 e.g. figure 2 in C07 suggests approximately ${\cal F} \propto M_H^{-0.3}$ if the mean field strength is of the order $5-6 ~\mu$G inside a radio halo of mass $M_H \sim 10^{14.5}$ M$_{\odot}$, assuming $B_H \propto M_H^{0.5}$. A shallower $B_H-M_H$ relation will correspondingly imply an weaker field to explain the observed $P_{1.4}-M_H$ scaling.

\vspace*{-2mm}
\section{Conclusions}

In this letter we presented the first radio-SZ correlation results for clusters hosting radio haloes, using published radio data and the \plk SZ catalog. There is a clear correspondence between these two thermal and non-thermal components as expected from the well-established radio/X-ray correlation.  
On the other hand, we found no strong bi-modal division in the cluster population split between radio halo and ``radio quiet'' objects. The halo non-detection clusters are generally radio under-luminous, but their occurrence in the \plk catalog is much less frequent as compared to all X-ray complete samples, and as such we can not conclude whether the bi-modality is weaker or non-existent when measured against SZ.  A likely explanation for this difference can be that the bi-modality seen in the $L_X$ selection comes from a bias towards lower mass cool core systems (which are radio quiet), whereas SZ selection picks up the most massive systems irrespective of their dynamical states. 
A forthcoming work will purport to test this hypothesis using a complete SZ-selected sample.

The radio-SZ correlation results were compared with the simplified theoretical predictions from hadronic and turbulent re-acceleration models. Even though the observed correlation power can be explained from both these models under certain assumptions, the turbulent re-acceleration model can be considered a better fit to the data given the simple formulations used in this letter. 
The difference between the global radio-SZ scaling and the one within the halo volume is explainable from the non-linear scaling between radio halo mass and total cluster mass. An indicative flattening of the correlation slope was observed when considering a cluster sample at 327 MHz, but the result became consistent with 1.4 GHz observations when the haloes and non-detections were considered together as one population.

\vspace*{-2mm}
\section*{Acknowledgments}
I am grateful to Klaus Dolag and Christoph Pfrommer for their help in clarifying the theory and observations of radio haloes. I  thank Arif Babul, Luigi Iapichino, Silvano Molendi, Florian Pacaud and Martin Sommer 
for helpful discussions, Mariachiara Rossetti for providing a temperature estimate for the cluster AS780, and in particular the anonymous referee for a thorough reading of the manuscript and suggesting numerous improvements. I acknowledge the invitation to participate in the KITP Program ``Galaxy clusters: the crossroads of astrophysics and cosmology'', supported by the NSF Grant No. PHY05-51164, where this research was initiated.

\label{lastpage}

\end{document}